# Entropy growth in the early universe and confirmation of initial big bang conditions
## (Why the quark-gluon model is not the best analogy)

### Andrew Beckwith
abeckwith@uh.edu, beckwith@aibep.org

**PACs:** 98.80.Cq, 95.35.+d, 95.30.Sf, 65.40.gd, 11.25.Uv

## Abstract
This paper shows how increased entropy values from an initially low big bang level can be measured experimentally by counting relic gravitons. Furthermore the physical mechanism of this entropy increase is explained via analogies with early-universe phase transitions. The role of Jack Ng's (2007, 2008a, 2008b) revised infinite quantum statistics in the physics of gravitational wave detection is acknowledged. Ng's infinite quantum statistics can be used to show that $\Delta S \approx \Delta N_{gravitons}$ is a starting point to the increasing net universe cosmological entropy. Finally, in a nod to similarities with ZPE analysis, it is important to note that the resulting $\Delta S \approx \Delta N_{gravitons} \neq 10^{88}$, that in fact it is much lower, allowing for evaluating initial graviton production as an emergent field phenomena, which may be similar to how ZPE states can be used to extract energy from a vacuum if entropy is not maximized. The rapid increase in entropy so alluded to without near sudden increases to $10^{88}$ may be enough to allow successful modeling of relic graviton production for entropy in a manner similar to ZPE energy extraction from a vacuum state. Doing so will provide clues as to how to re configure the Li-Baker detector to obtain relic graviton data sets from moments at or near the initial phases of the big bang.

## INTRODUCTION

This paper has the following chapter outline which will indicate what models of entropy may work. In addition, a de facto caution as to why string theory models may break down at the cosmic singularity is alluded to. In order to start off this analysis, we begin with the following chapters , one after the other.

1) Limitations of the Quark-Gluon analogy and how such limitations impact AdS/CFT correspondence applications 2) Ng's infinite quantum statistics. Is there a linkage of DM and Gravitons? 3) Quantum gas and applications of Wheeler De Witt equation to forming Partition function 4) Brane-antibrane 'pairs' and a linkage to Ng's quantum infinite statistics ? 5) Entropy, comparing values from T(u,v) stress energy , black holes, and general entropy values obtainable for the universe 6) Seth Lloyd's incomplete hypothesis 7) Simple relationships to consider (with regards to equivalence relation ships used to evaluate T(u,v)) 8) Data compression, continuity, and Dowker's space time sorting algorithm 9) Controversies of DM/ DE applications to cosmology. How HFGW may help resolve them.

A proper analysis of the onset of entropy growth may be crucial in resolving the role of DM/ DE to de facto cosmological evolution and the formation of early structure after the big bang. In addition we assert that if or not the Milky Way galaxy is in a void may be only resolvable via HFGW, for reasons which will be explained in the manuscript.

### Does "entropy" have an explicit meaning in astrophysics?

This paper will assert that there is a possibility of an equivalence between predicted Wheeler De Witt equation early universe conditions and the methodology of string theory, based upon a possible relationship between a counting algorithm for predicting entropy, based upon an article by Jack Ng (which he cites string theory as a way to derive his counting algorithm for entropy). This is due to re stating as entropy $S \approx <n> \big|_{gravitons}$ with $<n>$ as a numerical graviton density   and the expression given by Glinka (2007) for entropy (where Glinka uses the Wheeler De Witt equation), if we identify $\Omega$ as a partition function due to a graviton-quintessence gas.  If confirmed, this may also lead to new ways to model gravity/ graviton generation as part of an emergent 'field' phenomenon. Now why would anyone wish to revisit this problem in the first place? The reason is because that there are doubts people understand entropy in the first place. As an example of present confusion, please consider the following discussion where leading cosmologists, i.e. Sean Carroll (2005) asserted that there is a distinct possibility that mega black holes in the center of spiral galaxies have more entropy, in a calculated sense, i.e. up to $10^{90}$ in non dimensional units. This has to be compared to Carroll's (2005) stated value of up to $10^{88}$ in non dimensional units for observable non dimensional entropy units for the observable universe. Assume that there are over one billion spiral galaxies, with massive black holes in their center, each with entropy $10^{90}$, and then there is due to spiral galaxy entropy contributions $10^6 \times 10^{90} = 10^{96}$ entropy units to contend with, vs. $10^{88}$ entropy units to contend with for the observed universe. I.e. at least a ten to the eight order difference in entropy magnitude to contend with. The author is convinced after trial and error that the standard which should be used is that of talking of information, in the Shannon sense, for entropy, and to find ways to make a relationship between quantum computing operations, and Shannon information. Making the identification of entropy as being written as $S \sim \ln[partition-function]$. This is Shannon information theory with regards to entropy, and the convention will be the core of this text. What is chosen as a partition function will vary with our chosen model of how to input energy into our present universe. This idea as to an input of energy, and picking different models of how to do so leading to partition functions models is what motivated research in entropy generation. From now on, there will be an effort made to identify different procedural representations of the partiton function, and the log of the partion function with both string theory representations, i.e. the particle count algorithm of Y.Jack Ng, and the Wheeler De Witt version of the log of the partition function as presented by Glinka (2007). Doing so may enable researchers to eventually determine if or not gravity/ gravitational waves are an emergent field phenomenon.

Let us now examine candidates for entropy and discuss their advantages and limitations.

## Cautions as to what NOT to do with entropy-data compression-ZPE

In this inquiry, we should take care not to fall into several pit falls of analysis. We should avoid conflating any conceivable connections of zero point energy extraction, especially ZPE, and fluctuation states of ZPE, with data compression. The two do not mix for reasons which will be elaborated upon in the text. Secondly, the discussion we are embarking upon has no connection with intelligent design. As has been noted by Amara Angelica, typical data compression involves thousands of lines of computer coding, and on the face of it, it is preposterous to assume in any shape or form that self organizing systems, as an example, would duplicate machine coding in terms of data 'stream lining'. Lossless data compression is a class of data compression algorithms that allows the exact original data to be reconstructed from the compressed data. If we are to, as an example, reconstruct information pertinent to keeping the same cosmological parameter values of $\hbar, G, \alpha,$ from a prior universe to our present, it is important to delineate physical processes allowing for lossless data compression, and to do it in a way which has connections with theologically tained arguments contaminated with Intelligent Design thinking.

Alan Heavens(2000) identified criteria in an arXIV article stating "We show that, if the noise in the data is independent of the parameters, we can form linear combinations of the data which contain as much information about all the parameters as the entire dataset, in the sense that the Fisher information matrices are identical", as a criteria for lossless data compression. For our purposes, this is similar to reducing noise in 'information transfer' from a prior to a present universe as an ignorable datum which has no bearing upon encoded information values of $\hbar, G, \alpha,$ from a prior universe to our present. So how do we make noise as inconsequential ?

We have reviewed A.K. Avessian's article (2009) about alleged time variation of Planck's constant from the early universe, and found the arguments incomplete, again for reasons which will be discussed later on in the text. What is important would be to specify what would be a minimal amount of information needed to encode/ transmit a value of $\hbar$ from cosmological cycle to cycle, and to concentrate upon a lossless 'data compression' physical process for transmission of that $\hbar$ from cosmological cycle to cycle. The main criteria will be to identify a physical process where 'noise' in the transmitted data is independent of the value of $\hbar$ from cosmological cycle to cycle. The preferred venue for such a transmission would be a worm hole bridge from a prior to the present universe, as specified by both A.W. Beckwith (2008) at STAIF, and Lawrence B. Crowell (2005). John Preskill in 1994 came up with in his Caltech theory talk "information and black holes " a criteria for the physics of information encoded in very small distances, but we are convinced that something closer to deformation mechanics treatment of quantization criteria will be necessary , as given by Danien Sternheimer (2000) to tell us how to transmit a value of $\hbar$ from cosmological cycle to cycle, via either a worm hole, or similar process, while maintaining lossless 'data compression' by making whatever noise would be transmitted as inconsequential to keeping basic parameters intact from cycle to cosmological cycle. The identification of information, as we are doing here , with entropy measurements, makes a through investigation of entropy, and its generation as of high consequence to this inquiry.

## Minimum amount of information needed to initiate placing values of fundamental cosmological parameters

A.K. Avessian's article (2009) about alleged time variation of Planck's constant from the early universe depends heavily upon initial starting points for $\hbar(t)$, as given below, where we pick our own values for the time parameters, for reasons we will justify in this manuscript:

$$\hbar(t) \equiv \hbar_{initial}\left[t_{initial} \le t_{Planck}\right] \cdot \exp\left[-H_{macro} \cdot (\Delta t \sim t_{Planck})\right] \qquad (1)$$

The idea is that we are assuming a granular , discrete nature of space time. Futhermore, after a time we will state as $t \sim t_{Planck}$ there is a transition to a present value of space time, which is then probably going to be held constant.

It is easy to, in this situation, to get an inter relationship of what $\hbar(t)$ is with respect to the other physical parameters , i.e. having the values of $\alpha$ written as $\alpha(t) = e^2/\hbar(t) \cdot c$, as well as note how little the fine structure constant actually varies . Note that if we assume an unchanging Planck's mass $m_{Planck} = \sqrt{\hbar(t)c/G(t)} \sim 1.2 \times 10^{19} GeV$, this means that G has a time variance, too.

This leads to us asking what can be done to get a starting value of $\hbar_{initial}\left[t_{initial} \le t_{Planck}\right]$ recycled from a prior universe, to our present universe value. What is the initial value, and how does one insure its existence?

We obtain a minimum value as far as 'information' via appealing to Hogans (2002) argument where we have a maximum entropy as

$$S_{max} = \pi/H^2 \qquad (2)$$

, and this can be compared with A.K. Avessian's article (2009) value of, where we pick $\Lambda \sim 1$

$$H_{macro} \equiv \Lambda \cdot [H_{Hubble} = H] \qquad (3)$$

I.e. a choice as to how $\hbar(t)$ has an initial value, and entropy as scale valued by $S_{max} = \pi/H^2$ gives us a ball park estimate as to compressed values of $\hbar_{initial}\left[t_{initial} \le t_{Planck}\right]$ which would be transferred from a prior universe, to todays universe. If $S_{max} = \pi/H^2 \sim 10^5$, this would mean an incredibly small value for the INITIAL $H$ parameter, i.e. in pre inflation, we would have practically NO increase in expansion, just before the introduction vacuum energy, or emergent field energy from a prior universe, to our present universe.

Typically though, the value of the Hubble parameter, during inflation itself is HUGE, i.e. H is many times larger than 1, leading to initially very small entropy values. This means that we have to assume, initially, for a minimum transfer of entropy/ information from a prior universe, that H is neligible. If we look at Hogan's holographic model, this is consistent with a non finite event horizon

$$r_0 = H^{-1} \qquad (4)$$

This is tied in with a temperature as given by

$$T_{black-hole} = (2\pi \cdot r_0)^{-1} \qquad (5)$$

Nearly infinite temperatures are associated with tiny event horizon values, which in turn are linked to huge Hubble parameters of expansion. Whereas initially nearly zero values of temperature can be arguably linked to nearly non existent H values, which in term would be consistent with $S_{max} = \pi/H^2 \sim 10^5$ as a starting point to entropy. We next then must consider how the values of initial entropy are linkable to other physical models. I.e. can there be a transfer of entropy/ information from a pre inflation state to the present universe. Doing this will require that we keep in mind, as Hogan writes, that the number of distinguishable states is writable as

$$N = \exp(\pi H^{-2}) \qquad (6)$$

If, in this situation, that N is proportional to entropy, i.e. N as ~ number of entropy states to consider, , then as H drops in size, as would happen in pre inflation conditions, we will have opportunities for N ~ $10^5$

## Is data compression a way to distinguish what information is transferred to the present universe ?

The peak temperature as recorded by Weinberg (1972) is of the order of $10^{32}$ Kelvin, and that would imply using the expansion parameter, H, as given by Eqn (5) above. Likely before the onset of inflation, due to dimensional arguments, it can be safe to call the pre inflation temperature, T as very low. I.e. there was a build up of temperature, T, at the instant before inflation, which peaked shortly afterwards. Such an eventuality would be consistent with use of a worm hole bridge from a prior to a present universe. Beckwith (2008) at STAIF used such a model as a transfer of energy to the present universe, using formalism from Lawrence Crowell's book (2005)

A useful model as far as rapid transfer of energy would likely be a quantum flux, as provided for in Deformation quantization. We will follow the following convention as far as initiating quantization, i.e. the reported idea of Weyl quantization which is as follows: For a classical $u(p,q)$, a corresponding quantum observable is definable via

$$\Omega(u(p.q)) = \int_{\Re}^{2l} \tilde{u}(\xi,\eta) \cdot \exp[i \cdot (p \cdot \xi + q \cdot \eta)/\hbar] \cdot w(\xi,\eta) \cdot d^l\xi \cdot d^l\eta \qquad (7)$$

Here, C is the inverse fourier transform, and w(,) is a weight function, and p, and q are canonical variables fitting into $[p_\alpha, q_\beta] = i\hbar \cdot \delta_{\alpha,\beta}$, and the integral is taken over weak topology. For a quantized procedure as far as refinement of poisson brackets, the above, Weyl quantization is , as noted by S. Gutt and S. Waldemann (2006) equivalent to finding an operation $\Omega$ for which we can write

$$\Omega(1) = id \qquad (8)$$

As well as for Poisson brackets, $\{u,v\}$, obeying $(d/dt)u = -\{H,u\}$, and $(i/h) \cdot [u,H] = du/dt$

$$[\Omega(u), \Omega(v)] = i\hbar\Omega(\{u,v\}) \qquad (9)$$

For very small regimes of spatial integration, we can approximate Eqn. (7) as a finite sum, with

$$\tilde{u}(\xi,\eta) \sim \Omega(u) \qquad (10)$$

What we are doing is to give the following numerical approximate value of , de facto, as follows

$$\Omega(u(p.q)) = \int_{\Re}^{2l} \tilde{u}(\xi,\eta) \cdot \exp[i \cdot (p \cdot \xi + q \cdot \eta)/\hbar] \cdot w(\xi,\eta) \cdot d^l\xi \cdot d^l\eta \propto \tilde{u}(\xi_{initial}, \eta_{initial})$$

, and then we can state that the inverse transform is a form of data compression of information . Here, we will state that $\tilde{u}(\xi,\eta) \sim \Omega(u) \sim$ {**information bits for** } $\hbar_{initial}[t_{initial} \leq t_{Planck}]$ as far as initial values of the plancks constant are concerned. Please see **Appendix 1** as to how for thin shell geometries the Weyl quantization condition reduces to the Wheeler De Witt equation. I.e. a wave functional approximately presentable as

$$\Psi \sim [R/R_{eq}]^{3/2} \qquad (10a)$$

, where R refers to a spatial distance from the center of a spherical universe. **Appendix II** is an accounting of what is known as a pseudo time dependent solution to the Wheeler de Witt equation involving a worm hole bridge between two universes. The metric assumed in Appendix I is a typical maximally symmetric metric, whereas Appendix 2 is using the Reisssner- Nordstrom metric. We assume, that to first order, if the value of R in $\Psi \sim [R/R_{eq}]^{3/2}$ is nearly $R \propto l_P \sim 10^{43}$ centimeters, I.e. close to singularity conditions, that the issue of how much information from a prior universe, to our own may be addressed, and that the solution $\Psi \sim [R/R_{eq}]^{3/2}$ is consistent with regards to Weyl geometry. So let us consider what information is transferred . We claim that it centers about enough information with regards to preserving $\hbar$ from universe cycle to cycle.

## Information bits for $\hbar_{initial}[t_{initial} \leq t_{Planck}]$ as far as initial values of the Plancks constant are concerned

To begin this inquiry, it is appropriate to note that we are assuming that there is a variation in the value of $\Psi \sim [R/R_{eq}]^{3/2}$ with a minimum value of $R \propto l_P \sim 10^{-43}$ centimeters to work with. Note that Honig's (1973) article specified a general value of about $3.68 \times 10^{-48}$ grams, per photon, and that each photon has an energy of $E[photon] = \frac{hc}{\lambda} = m_{photon} \cdot c^2$. If one photon is, in energy equivalent to $10^{12}$ gravitons, then, if $\lambda \sim l_P$ = Planck's length, gives us a flux value as to how many gravitons / entropy units are transmited. The key point is that we wish to determine what is a minimum amount of information bits/attendant entropy values needed for transmission of $\hbar_{initial}[t_{initial} \leq t_{Planck}]$ . In order to do this, note the article, i.e. a "A minimum photon "rest mass" — Using Planck's constant and discontinuous electromagnetic waves which as written in September, 1974 *by William Honig specifies a photon rest mass of the order of* $3.68 \times 10^{-48}$ *grams per photon.*

If we specify a mass of about $10^{-60}$ grams per graviton, then to get at least one photon, and if we use photons as a way of 'encapsulating' $\hbar_{initial}[t_{initial} \leq t_{Planck}]$, then to first order, we need about $10^{12}$ gravitons / entropy units ( each graviton, in the beginning being designated as one 'carrier container' of information for one unit of $\hbar_{initial}[t_{initial} \leq t_{Planck}]$. If as an example, as calculated by Beckwith (2008) that there were about $10^{21}$ gravitons introduced during the onset of inflation , this means a minimum copy of about one billion $\hbar_{initial}[t_{initial} \leq t_{Planck}]$ information packets being introduced from a prior universe, to our present universe, i.e. more than enough to insure introducing enough copies of $\hbar_{initial}[t_{initial} \leq t_{Planck}]$ to insure continuity of physical processes.

For those who doubt that $10^{-60}$ grams per graviton can be reconciled with observational tests with respect to the Equivalence Principle and all classical weak-field tests , we refer the readers to Matt Visser's (1998) article about "Mass for the graviton". The heart of Matt Visser's calculation for a non zero graviton mass involve placing appropriate small off diagnoal terms to the usual stress tensor T (u,v) calculation, a development which in certain ways fore shadows what was done by C.S. Unnikrishnan's (2009) revisement of special relativity, in ways which will be described in this document.

## Limitations of the Quark-Gluon analogy and how such limitations impact AdS/CFT correspondence applications

What is being alluded to, is that variations in the AdS/CFT correspondence applications exist from what is usually assumed for usual matter. The differences, which are due to quark-gluon plasma models breaking down in the beginning of the big bang point to the necessity of using something similar to the counting algorithm as introduced by Ng, as a replacement for typical string theory models in strict accordance to AdS/CFT correspondence.

The goal of exploring the degree of divergence from AdS/CFT correspondence will be in quantifying a time sequence in evolution of the big bang where there is a break from causal continuity.

A break down in causal continuity will, if confirmed, be a way of signifying that encoded information from a prior era has to be passed through to the present universe in likely an emergent field configuration.. If much of the information is passed to our present universe in an emergent field configuration , this leaves open the question of if or not there is a time sequence right after the initial phases of the big bang where there was a re constitution of information in traditional four space geometry.

One candidate for specifying such a re constitution of entropy / space time information would be to model gravitons as kink – anti kink 'pairs' which are re constituted in space time right after the big bang. Conceivably, in such a situation, a fifth 'higher dimension' could be a conduit of 'graviton' information / entropy packaging from a prior universe, to our own, with the spill over of this information being re constituted in four space with the re appearance of , via kink – anti kink pairs after a thermal phase transition occurs.

The problem with implying data is compressed, is that this, at least by popular imagination implies highly specific machine / IT analogues. We wish to assure the readers that no such appeal to intelligent design/ deity based arguments is implied in this document.

A point where there is a breakage in causal continuity will help determine if or not there is a reason for data compression. In computer science and information theory, data compression or source coding is the process of encoding information using fewer bits (or other information-bearing units) than an unencoded representation would use through use of specific encoding schemes. Using fewer bits of an encoding scheme for 'information' may in its own way allow data compression. We need to have a similar model for explaining the degree of information transferred from a prior universe, to the present, while maintaining the structural integrity of the basic cosmological parameters, such as $\hbar$, G, and the fine structure constant. $\alpha$. In doing so , we will make the identification of information, with entropy, in effect mimicking simple entropy coding for infinite input data with a geometric distribution.

Witten, Radford M. Neal, and John G. Cleary, *Arithmetic Coding for Data Compression*, CACM 30(6):520-540, June 1987 write that the very best compressors use probabilistic models which predictions are coupled to an algorithm called arithmetic coding. Again, while avoiding the intelligent design analogies, it is possible to imply that if there was a restriction of information to dimensions other than the typical space time dimensions of four space, with fifth and higher dimensions being our information conduit, that by default, data compression did occur during the restriction of much of the information incoded in kink- anti kink gravitons disappearing before the big bang in four space, and then re appearing in our present day four space geometry, as a spill over from a fifth dimension

Since there is a problem physicists, writers, and editors have with any remote degree of ambiguity, let us briefly review what is known about singularity theorems for GR, in four space. Then make a reasonable extrapolation in fifth space embedding of the four dimensions, to make our point about singularities in four space more understandable

A. Feinstein, et al, (2000) laid out how singularities can be removed from the higher-dimensional model when only one of the extra dimensions is time-varying. If the fifth dimension has , indeed time variance, in any number of ways, four dimensional singularities no longer have the same impact with a time dependent fifth dimensions. And a varying fifth higher dimension is, in itself, a perfect conduit for information from a prior universe, to our present universe. The information restriction in four dimensions, then in four dimensions is a causal discontinuity, while the fifth dimension, with its time variance will be, due to information restriction from four dimensions, our avenue for data 'compression' of transferral of information from a prior universe to our present. Furthermore, the spill over from a restriction in four space, to five spaces, with the dumping of information/ entropy in present four space after it transfers from a prior universe corresponds to graviton re combination from kink- anti kink structures as entropy increases from a very low level.

This will be done, especially when entropy is held to be in tandem with a 'particle count' of instanton-anti instanton packaged gravitons as the mechanism for increase of entropy from a much lower level to todays level .

If the instanton-anti instanton portrayal of gravitons is held to be legitimate, and if the kink- anti kink structure of gravitons materializes as part of an emergent field from a near singularity at the onset of inflation, then the question of data compression becomes urgent. I.e. what lead to the nucleation of information in a kink-anti kink configuration. Does this configuration have ANY commonality with what was placed into the singularity-near singularity at the nextus of a prior universe' collapse. In addition, this also raises the possibility of the Penrose alternative to a typical cyclic universe. To re count what Penrose suggested, he used typical inflaton expansion from a singular starting point, i.e. through the D'Albertain equation, but then assumed that the universe continued expanding forever. His model specifically alluded to black holes collating matter-energy, with an unsaid , unspecified 'mapping ' function of sucked up matter-energy in each black hole and then rcycling matter-energy so collated from millions of black holes to a new starting point. This was presented in Penn state, (2007). An advantage of appropriate encryption scheme from a prior universe to the present may lie in helping to eventually obtaining falsifiable experimental criteria to distinguish between the Penrose suggestion, as outlined above, and more typical big bang- big crunch models . i.e. the Steinhart model (2002)

To begin this analysis, let us look at what goes wrong in models of the early universe. The assertion made is that this is due to the quark – Gluon model of plasmas having major 'counting algorithm' breaks with non counting algorithm conditions, i.e. when plasma physics conditions BEFORE the advent of the Quark gluon plasma existed. Here are some questions which need to be asked.

1. Is QGP strongly coupled or not? Note : Strong coupling is a natural explanation for the small (viscosity)
 Analogy to the RHIC: J/y survives deconfinement phase transition
2. What is the nature of viscosity in the early universe? What is the standard story?  (Hint: AdS-CFT correspondence models). Question 2 comes up since

$$\frac{\eta}{s} = \frac{1}{4\pi} \quad (11)$$

 typically holds for liquid helium and most bosonic matter. However, this relation breaks down. At the beginning of the big bang. As follows i.e. if Gauss- Bonet gravity is assumed, in order to still keep casuality , one needs $\lambda_{BG} \leq \frac{9}{100}$

This even if one writes for a viscosity over entropy ratio the following

$$\frac{\eta}{s} \equiv \frac{1}{4\pi} \cdot [1 - 4\lambda_{GB}] \leq \frac{1}{4\pi} \tag{12}$$

A careful researcher may ask why this is so important. If a causal discontinuity as indicated means the $\frac{\eta}{s}$ ratio is $\approx \frac{1}{4\pi} \cdot \frac{33}{50}$, or less in value, it puts major restrictions upon viscosity, as well as entropy. A drop in viscosity, which can lead to major deviations from $\frac{1}{4\pi}$ in typical models may be due to more collisions. Then, more collisions due to WHAT physical process? Recall the argument put up earlier. I.e. the reference to causal discontinuity in four dimensions, and a restriction of information flow to a fifth dimension at the onset of the big bang/ transition from a prior universe? That process of a collision increase may be inherent in the restriction to a fifth dimension, just before the big bang singularity, in four dimensions, of information flow. In fact, it very well be true, that initially, during the process of restriction to a 5$^{th}$ dimension, right before the big bang, that $\left|\frac{\eta}{s} \approx \varepsilon^+\right| << \frac{1}{4\pi}$. Either the viscosity drops nearly to zero, or else the entropy density may, partly due to restriction in geometric 'sizing' may become effectively nearly infinite

It is due to the following qualifications put in about Quark – Gluon plasmas which will be put up, here. **Namely,** more collisions imply less viscosity. More Deflections ALSO implies less viscosity. Finally, the more momentum transport is prevented, the less the viscosity value becomes. Say that a physics researcher is looking at viscosity due to turbulent fields. Also, perturbatively calculated viscosities: due to collisions. This has been known as *Anomalous Viscosity* in plasma physics ,(this is going nowhere, from pre-big bang to big bang cosmology).

So happens that **RHIC models for viscosity assume**

$$\frac{1}{\eta} \approx \frac{1}{\eta_A} + \frac{1}{\eta_C} \tag{13}$$

As Akazawa noted in an RHIC study, equation 13 above makes sense if one has stable temperature T, so that

$$\frac{\eta_A}{s} = \bar{c}_0 \cdot \left(\frac{T}{g^2 |\nabla u|}\right)^{\frac{2n-1}{2n+1}} \Leftrightarrow \frac{\eta_C}{s} = \text{constant} \tag{14}$$

If the temperature T wildly varies, as it does at the onset of the big bang, this breaks down completely. This development is **Mission impossible: why we need a different argument for entropy. I.e. Even for the RHIC, and in computational models of the viscosity for closed geometries—what goes wrong in computational models**
- **Viscous Stress is *NOT* $\propto$ shear**
- **Nonlinear response: impossible to obtain on lattice ( computationally speaking)**
- **Bottom line: we DO NOT have a way to even define SHEAR in the vicinity of big bang!!!!**

We now need to ask ourselves what may be a way to present entropy/ entropy density in a manner which may be consistent with having / explaining how $\left|\frac{\eta}{s} \approx \varepsilon^+\right| << \frac{1}{4\pi}$ may occur, and also what may be necessary to explain how the entropy / entropy density may become extraordinarily large, and that, outside of the restriction to a fifth dimension argument mentioned earlier for 'information' transferral to the onset of the big bang, that it is not necessary to appeal to nearly infinite collisions in order to have a drop in viscosity. This will lead us to Y. Jack Ng's 'particle count' algorithm for entropy, below

## Is each 'particle count unit' as brought up by Ng, is equivalent to a brane-anti brane 'unit in brane treatments of entropy?

It is useful to state this convention for analyzing the resulting entropy calculations, because it is a way to explain how and why the number of instanton – anti instanton pairs, and their formulation and break up can be linked to the growth of entropy. If, as an example, there is a linkage between quantum energy level components of the quantum gas as brought up by Glinka (2007) and the number of instanton- anti instanton pairs, then it is possible to ascertain a linkage between a Wheeler De Witt worm hole introduction of vacuum energy from a prior universe to our present universe, and the resulting brane- anti brane (instanton-anti instanton) units of entropy. What would be ideal would be to make an equivalence between a quantum number, n, say of a quantum graviton gas, as entering a worm hole, i.e. going back to the Energy ( quantum gas ) $\approx n \cdot \hbar \omega$, and the number <n> of pairs of brane- anti brane pairs showing up in an entropy count, and the growth of entropy. We are fortunate that Dr. Jack Ng's research into entropy ( 2007,2008) not only used the Shannon entropy model, but also as part of his quantum infinite statistics lead to a quantum counting algorithm with entropy proportional to 'emergent field' particles. If as an example a quantum graviton gas exists, as suggested by Glinka(2007) , if each quantum gas 'particle' is equivalent to a graviton, and that graviton is an 'emergent' from quantum vacuum entity, then we fortuitously connect our research with gravitons with Shannon entropy, as given by $S \sim \ln[partition - function]$. This is a counter part as to what *Asakawa* et al, (2001, 2006) suggested for quark gluon gases, and the 2$^{nd}$ order phase transition written up by Torrieri et al (2008) brought up at the nuclear physics Erice (2008) school, in discussions with the author.

Furthermore, finding out if or not it is either a drop in viscosity, when $\left|\frac{\eta}{s} \approx \varepsilon^+\right| << \frac{1}{4\pi}$, or a major increase in entropy density may tell us how much information is , indeed, transferred from a prior universe to our present. If it is $s \to \infty$, for all effective purposes, at the moment after the pre big bang configuration , likely then there will be a high degree of 'information' from a prior universe exchanged to our present universe. If on the other hand, $\eta \to 0^+$ due to restriction of 'information from four dimensional 'geometry' to a variable fifth dimension, so as to indicate almost infinite collisions with a closure of a fourth dimensional 'portal' for information flow, then it is likely that significant data compression has occurred. While stating this, it is note worthy to state that the Penrose-Hawking singularity theorems do not give precise answers as to information flow from a prior to the present universe. Hawking's singularity theorem is for the whole universe, and works backwards-in-time: it guarantees that the big-bang has infinite density. This theorem is more restricted, it only holds when matter obeys a stronger energy condition, called the *dominant energy condition*, which means that the energy is bigger than the pressure. All ordinary matter, with the exception of a vacuum expectation value of a scalar field, obeys this condition.

This leaves open the question of if or not there is 'infinite' density of ordinary matter, or if or not there is a fifth dimensional leakage of 'information' from a prior universe to our present. If there is merely infinite 'density', and possibly infinite entropy 'density/ disorder at the origin, then perhaps no information from a prior universe is transferred to our present universe.On the other hand, having $\eta \to 0^+$, or at least be very small may indicate that data compression is a de rigor way of treating how information for cosmological parameters, such as $\hbar$, G, and the fine structure constant. $\alpha$ arose, and may have been recycled from a prior universe..

Details about this have to be worked out, and this because that as of present one of the few tools which is left to formulation and proof of the singularity theorems is the Raychaudhuri equation, which describes the divergence θ of a congruence (family) of geodesics, which has a lot of assumptions behind it, as stated by [Naresh Dadhich](#)(2005). As indicated by Hawkings theorem, infinite density is its usual modus operandi, for a singularity, and this assumption may have to be revisited. Natário, J. (2006) has more details on the different type of singularities involved.

## Ng's particle count algorithm and Brane models of entropy

Let us first summarize what can be said about Ng's quantum infinite statistics. Afterwards, the numerical counting involved has a direct connection with the pairs of brane- anti brane (kink- anti kink) structures Mathur (2007) and others worked with to get an entropy count. Ng (2007, 2008) outlines how to get $S \approx N$, which with additional arguments we refine to be $S \approx <n>$ (where <n> is a 'DM' density). Begin with a partition function. As given by Ng (2007,2008) and referenced by Beckwith (2009)

$$Z_N \sim \left(\frac{1}{N!}\right) \cdot \left(\frac{V}{\lambda^3}\right)^N \tag{15}$$

This, according to Ng, leads to entropy of the limiting value of

$$S \approx N \cdot \left(\log\left[V/N\lambda^3\right] + 5/2\right) \tag{16}$$

But $V \approx R_H^3 \approx \lambda^3$, so unless N in Eqn (16) above is about 1, S (entropy) would be < 0, which is a contradiction. Now this is where Jack Ng introduces removing the N! term in Eqn (15) above, i.e., inside the Log expression we remove the expression of N in Eqn. (16) above. This is a way to obtain what Ng refers to as Quantum Boltzmann statistics, so then we obtain for sufficiently large N

$$S \approx N \tag{17}$$

The supposition is that the value of N is proportional to a numerical DM density referred to as $<n> \big|_{Dark-matter}$. HFGW would play a role if $V \approx R_H^3 \approx \lambda^3$ has each $\lambda$ of the order of being within an order of magnitude of the Planck length value, as implied by Beckwith (2009). What the author is examining is, if or not there can be linkage made between $S \approx <n> \big|_{gravitons}$ and the expression given by Glinka (2007) of, if we identify $\Omega = \frac{1}{2|u|^2 - 1}$ as a partition function (with $u$ part of a Bogoliubov transformation) due to a graviton-quintessence gas, to get information theory based entropy

$$S \equiv \ln \Omega \tag{18}$$

Such a linkage would open up the possibility that the density of primordial gravitational waves could be examined, and linked to modeling gravity as an effective theory, as well as giving credence to how to avoid dS/dt = ∞ at S=0 . If so, then one can look at the research results of Samir Mathur (2007). This is part of what has been developed in the case of massless radiation, where for D space-time dimensions, and E, the general energy is

$$S \sim E^{(D-1/D)} \tag{19}$$

This suggests that entropy scaling is proportional to a power of the vacuum energy, i.e., entropy ~ vacuum energy, if $E \sim E_{total}$ is interpreted as a total net energy proportional to vacuum energy, as given below**.** Conventional brane theory actually enables this instanton structure analysis, as can be seen in the following. This is adapted from a lecture given at the ICGC-07 conference by Andrew Beckwith (2007b)

$$\frac{\Lambda_{Max} V_4}{8 \cdot \pi \cdot G} \sim T^{00} V_4 \equiv \rho \cdot V_4 = E_{total} \tag{20}$$

Traditionally, minimum length for space-time benchmarking has been via the quantum gravity modification of a minimum Planck length for a grid of space-time of Planck length, whereas this grid is changed to something bigger $l_P \sim 10^{-33} cm \xrightarrow{Quantum-Gravity-threshold} \widetilde{N}^\alpha \cdot l_P$. So far, we this only covers a typical string gas model for entropy. $\widetilde{N}$ is assigned as the as numerical density of brains and anti-branes. A brane-antibrane pair corresponds to solitons and anti-solitons in density wave physics. The branes are equivalent to instanton kinks in density wave physics, whereas the antibranes are an antiinstanton structure. First, a similar pairing in both black hole models and models of the early universe is examined, and a counting regime for the number of instanton and anti-instanton structures in both black holes and in early universe models is employed as a way to get a net entropy-information count value. One can observe this in the work of Gilad Lifschytz in 2004. Lifschyztz (2004) codified thermalization equations of the black hole, which were recovered from the model of branes and antibranes and a contribution to total vacuum energy. In lieu of assuming an antibrane is merely the charge conjugate of say a Dp brane. Here, $M_{p\,j,0}$ is the number of branes in an early universe configuration, while $M_{\bar{p}\,j,0}$ is antibrane number. I.e., there is a kink in the given $brane \sim M_{p\,j,0} \leftrightarrow CDW\ e^-$ electron charge and for the corresponding anti-kink $anti-brane \sim M_{\bar{p}\,j,0} \leftrightarrow CDW\ e^+$ positron charge. Here, in the bottom expression, $\breve{N}$ is the number of kink-anti-kink charge pairs, which is analogous to the simpler CDW structure.

$$S_{Total} \sim \breve{a} \cdot \left[\frac{E_{Total}}{2^n}\right]^\lambda \cdot \prod_{j=1}^{\breve{N}} \left(\sqrt{M_{p\,j,0}} + \sqrt{M_{\bar{p}\,j,0}}\right) \qquad (21)$$

This expression for entropy (based on the number of brane-anti-brane pairs) has a net energy value of $E_{Total}$ as expressed in Eqn (20) above, where $E_{Total}$ is proportional to the cosmological vacuum energy parameter; in string theory, $E_{Total}$ is also defined via

$$E_{Total} = 4\lambda \cdot \sqrt{M_{p\,j,0} \cdot M_{\bar{p}\,j,0}} \qquad (22)$$

Equation 12 can be changed and rescaled to treating the mass and the energy of the brane contribution along the lines of Mathur's CQG article (2007) where he has a string winding interpretation of energy: putting as much energy $E$ into string windings as possible via $[n_1 + \bar{n}_1]LT = [2n_1]LT = E/2$, where there are $n_1$ wrappings of a string about a cycle of the torus, and $\bar{n}_1$ being "wrappings the other way,", with the torus having a cycle of length $L$, which leads to an entropy defined in terms of an energy value of mass of $m_i = T_P \prod L_j$ ( $T_P$ is the tension of the $i$ th brane, and $L_j$ are spatial dimensions of a complex torus structure). The toroidal structure is to first approximation equivalent dimensionally to the minimum effective length of $\widetilde{N}^\alpha \cdot l_P \sim \widetilde{N}^\alpha$ times Planck length $\propto 10^{-35}$ centimeters

$$E_{Total} = 2\sum_i m_i n_i \qquad (23)$$

The windings of a string are given by figure 6.1 of Becker et al, as the number of times the strings wrap about a circle midway in the length of a cylinder. The structure the string wraps about is a compact object construct Dp branes and anti-branes. Compactness is used to roughly represent early universe conditions, and the brane-anti brane pairs are equivalent to a bit of "information.". This leads to entropy expressed as a strict numerical count of different pairs of Dp brane-Dp anti-branes, which form a higher-dimensional equivalent to graviton production. The tie in between Eqn. (24) below and Jack Ng's treatment of the growth of entropy is as follows: First, look at the expression below, which has $\breve{N}$ as a stated number of pairs of Dp brane-antibrane pairs: The suffix $\breve{N}$ is in a 1-1 relationship with $\Delta S \approx \Delta N_{gravitons}$

$$S_{Total} = A \cdot \prod_i^{\bar{N}} \sqrt{n_i} \qquad (24)$$

# Entropy, comparing values from T(u,v) stress energy, black holes, and general entropy values obtainable for the universe

We start off with looking at Vacuum energy and entropy. This suggests that entropy scaling is proportional to a power of the vacuum energy, i.e., entropy ~ vacuum energy, if is interpreted as a total net energy proportional to vacuum energy, i.e. go to equation 10 above. What will be done is hopefully, with proper analysis of T(u,v) at the onset of creation, is to distinguish, between entropy say of what Mathur wrote, as $S \sim E^{(D-1/D)}$, and see how it compares with the entropy of the center of the galaxy, as given by equation 25, as opposed to the entropy of the universe, as given by equation 16 below. The entropy which will be part of the resulting vacuum energy will be writable as either Black hole entropy and / or the Universe's entropy. I.e. for black hole entropy, from Sean Carroll (2005), the entropy of a huge black hole of mass M at the center of the milky way galaxy. Note there are at least a BILLION GALAXIES, and M is ENORMOUS

$$S_{Black-Hole} \sim 10^{90} \cdot \left[\frac{M}{10^6 \cdot M_{Solar-Mass}}\right]^2 \qquad (25)$$

This needs to be compared with the entropy of the universe, as given by Sean Carroll, as stated by

$$S_{Total} \sim 10^{88} \qquad (26)$$

The claim made here is that if one knew how to evaluate T(u,v) properly, that the up to $10^9$ difference in equations 25 and 26 will be understandable, and that what seems to be dealt with directly. Doing so is doable if one understands the difference/ similarities in equations 21, 23, and 24, above. So, how does one do this ? The candidate picked which may be able to obtain some commonality in the different entropy formalisms is to confront what is both right and wrong in Seth Lloyd's entropy treatment in terms of operations as given below. Furthermore, what is done should avoid the catastrophe inherent in solving the problem which Mithras gave the author, that of dS/dt =∞ at S=0 in Kochi, India, as a fault of classical GR which should be avoided. One of the main ways to perhaps solve this will be to pay attention to what C. S. Unnikrishnan put up in 2009, i.e. his article about the purported one way speed of light, and its impact upon perhaps a restatement of T(u,v). A re statement of how to evaluate T(u,v) may permit a proper frame of reference to close the gap between entropy values as given in Equations 25 and 26 above.

## Seth Lloyd's linking of information to entropy

By necessity, entropy will be examined, using the equivalence between number of operations which Seth Lloyd used in his model, and total units of entropy as the author referenced from Sean Carroll, and other theorists. The key equation Seth Lloyd wrote is as follows, assuming a low entropy value in the beginning

$$|S_{Total}| \sim |k_B \cdot \ln 2| \cdot [\# operations]^{3/4} \qquad (27)$$

Seth Lloyd is making a direct reference to a linkage between the number of operations a quantum computer model of how the Universe evolves is responsible for , in the onset of a big bang picture, and entropy..If equation 27 is accepted, which is debatable, then the issue is what is the unit of operation, i.e. the mechanism involved for an operation for assembling a graviton, and can that be reconciled with T(0,0) above. A good question is, if this is done, then how to get an appropriate operation, linkable with the number of emergent gravitons, so **at least** equation 27 will be congruent with

## Simple relationships to consider (with regards to equivalence relation ships used to evaluate T(u,v))

What needs to be understood and evaluated is, if there is a re structuring of an appropriate frame of reference for T(u,v) and its resultant effects upon how to reconcile black hole entropy, a.k.a. Eqn (25) with Eqn (26) and Eqn (27). A good place to start would be to obtain T(u,v) values which are consistent with slides on the two way versus one way light speed presentation of the ISEG 2009 conference. We wish to obtain T(u,v) values properly analyzed with respect to early universe metrics, and PROPERLY extrapolated to today so that ZPE energy extraction, as pursued by many, will be the model for an emergent field development of entropy. Note the easiest version of T(u,v) as presented by Wald . If metric g(a,b) is for curved space time, the simplest matter energy stress tensor is ( Klein Gordon)

$$T_{ab} = \nabla_a \phi \cdot \nabla_b \phi - \frac{1}{2} \cdot g_{ab} \cdot (\nabla_c \phi \cdot \nabla^c \phi + m^2 \phi) \qquad (28)$$

What is affected by Unnikrishnan's presented (2009) hypothesis is how to keep g(a,b) properly linked observationally to a Machian universe frame of reference, not the discredited aether, via CMBR spectra behavior. If the above equation is held to be appropriate, and then elaborated upon, the developed T(u,v) expression should adhere to Wald's unitary equivalence principle. **The structure of unitary equivalence is foundational to space time maps, and Wald states it as being**

$$C\mu_1(\psi,\psi) \leq \mu_2(\psi,\psi) \leq C'\mu_1(\psi,\psi) \qquad (29)$$

**While stating this, it is important to keep in mind that Wald defines**

$$\mu(\psi_1,\psi_2) = \operatorname{Im} \widetilde{\Omega}(K\psi_1, K\psi_2) = \frac{1}{2} \cdot <\psi_1, |A| \cdot \psi_2> \qquad (30)$$

We defined the operation, where A is a bounded operator, and < > an inner product via use of

$$<\psi_1, |A| \cdot \psi_2> = \int_\Sigma [T_{ab}(\psi_1, |A|\psi_2)] \xi^a \eta^b \sqrt{h} \cdot d^3 x \qquad (31)$$

The job will be to keep this same equivalence relationship intact for space time, no matter what is done with the metric g(a,b) in the T(a,b) expressions we work with, which will be elaborations of Eqn (28) above.

## Data compression, continuity, and Dowker's space time sorting algorithm

**Regarding information theory:**

This is closely tied in with data compression and how much 'information' material from a prior universe is transferred to our present universe. In order to do such an analysis of data compression and what is sent to out present universe from a prior universe, it is useful to consider how there would be an eventual increase in information / entropy terms, from $10^{21}$ to $10^{88}$. Too much rapid increase would lead to the same problem ZPE researchers have . I.e. if Entropy is maximized too quickly, we have no chance of extracting ZPE energy from a vacuum state, i.e. no emergent phenomena is possible. What to avoid is akin to avoiding

$$S_{gw} = V \cdot \int_{v_0}^{v_1} r(v) \cdot v^2 dv \cong (10^{29})^3 \cdot (H_1/M_P)^{3/2} \approx 10^{87} - 10^{88} \qquad (32)$$

Eqn (22) is from Giovanni, and it states that all entropy in the universe is solely due to graviton production. This absurd conclusion would be akin, in present day parlance, to having $10^{88}$ entropy 'units' created right at the onset of the big bang. This does NOT happen.

What will eventually need to be explained will be if or not $10^7$ entropy units, as information transferred from a prior big bang to our present universe would be enough to preserve $\hbar$, G, and other physical values from a prior universe, to today's present cosmology. Inevitably, if $10^7$ entropy/ information units are exchanged via data compression from a prior to our present universe, Eqn (27) , and resultant increases in entropy up to $10^{88}$ entropy 'units' will involve the singularity theorems of cosmology, as well as explanations as to how $\Delta S|_{relic-HFGW} \approx \Delta N \sim 10^{21}$ could take place, say right at the end of the inflationary era. The author claims that to do so, that Eqn (27) , and a mechanism for the assembly of gravitons from a kink- anti kink structure is a de rigor development. We need to find a way to experimentally verify this tally of results. And to find conditions under which the abrupt reformulation of a near-constant cosmological constant, i.e., more stable vacuum energy conditions right after the big bang itself, would allow for reformulation of SO(4) gauge-theory conditions.

**What is the bridge between low entropy of the early universe and its rapid build up later?** Penrose in a contribution to a conference, (2006) on page two of the Penrose conference (2006) document refers to the necessity of reconciling a tiny initial starting entropy of the beginnings of the universe with a much larger increased value of entropy substancially later. As can be read from the article by Penrose (2006) "A seeming paradox arises from the fact that our best evidence for the existence of the big bang arises from observations of the microwave background radiation-"….. " This corresponds to maximum entropy so we reasonably ask: how can this be consistent with the Second law, according to which the universe started with a tiny amount of entropy" . Penrose then goes on to state that " The answer lies in the fact that the high entropy of the microwave background only refers to the matter content of the universe, and not the gravitational field, as would be enclosed by its space-time background in accordance to Einstein's theory of general relativity". Penrose then goes on to state that the initial pre red shift equals 1100 background would be remarkably homogeneous. I.e. for red shift values far greater than 1100 the more homogeneous the universe would become according to the dictum that " gravitational degrees of freedom would not be excited at all" Beckwith (2008) then asks the question of how much of a contribution the baryonic matter contribution would be expected to make to entropy production. . The question should be asked in terms of the time line as to how the universe evolved, as specified by both Steinhardt and Turok (2007) on pages 20-21 of their book, as well as by NASA . And a way to start this would be to delineate further the amplitude vs frequency GW plot as given below. It is asserted that the presence of the peak in gravity wave frequency at about $10^{10}$ Hertz (shown in figure 1) has significant consequences for observational cosmology.

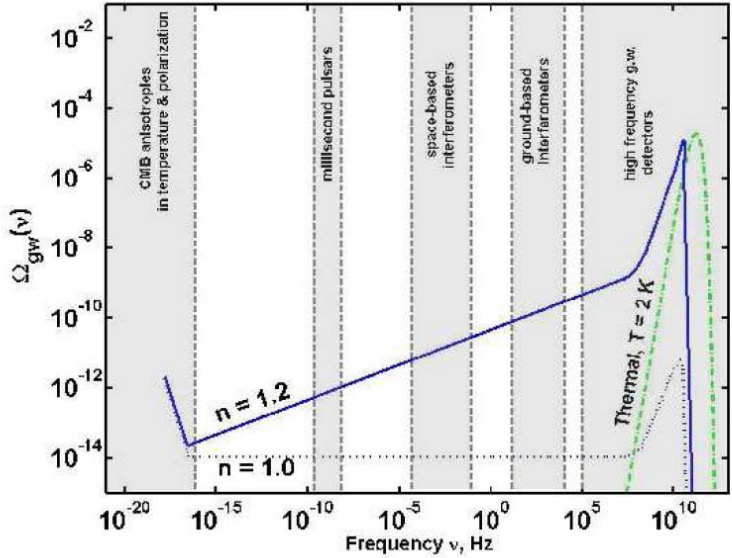

**FIGURE 1** Where HFGWs come from: Grishchuk found the maximum energy density (at a peak frequency) of relic gravitational waves (Grishchuk, 2007).

Finding an appropriate phase transition argument for the onset of entropy creation and graviton production

$$s_{Density} = \frac{2 \cdot \pi^2}{45} \cdot g_* \cdot T^3 \qquad (33)$$

is akin to explaining how, and why temperature changes in T, lead to , if the temperature increases, an emergent field description of how gravitons arose. We claim that this is identical to obtaining a physically consistent description of entropy density would be akin to, with increasing , then decreasing temperatures a study as to how kink- anti kink structure of gravitons developed . This would entail developing a consistent picture , via SO(4) theory of gravitons being assembled from a vacuum energy back ground and giving definition as to Seth Lloyd's computation operation description of entropy. Having said this, it is now appropriate to raise what gravitons/ HFGW may tell us about structural evolution issues in today's cosmology. Here are several issues the author is aware of which may be answered by judicious use of HFGWs.

As summarized by Thanu Padmanabhan (IUCAA) in the recent 25th IAGRG presentation he made, "Gravity: The Inside Story ", entropy can be thought of as due to 'ignored' degrees of freedom, classically, and is generalized in general relativity by appealing to to extremising entropy for all the null surfaces of space time. Padmanabhan claims the process of extemizing entropy then leads to equations for the background metric of the space-time. I .e. that the process of entropy being put in an entremal form leads to the Einsteinian equations of motion. What is done in this present work is more modest. I.e. entropy is thought of in terms of being increased by relic graviton production, and the discussion then examines the consequence of doing that in terms of GR space time metric evolution. How entropy production is tied in with graviton production is via recent work by Jack Ng. It would be exciting if or not we learn enough about entropy to determine if or not we can identify null surfaces, as Thanu brought up in his presentation in his Calcutta (2009). presentation. The venue of research brought up here we think is a step in just that direction. Furthermore, let us now look at large scale structural issues which may necessitate use of HFGW to resolve. Job one will be to explain what may the origins of the enormous energy spike in Figure 1 above, by paying attention to Relic gravitational waves , allowing us to make direct inferences about the early universe Hubble parameter and scale factor ("birth" of the Universe and its early dynamical evolution). According to Grishchuk: energy density requires that the GW frequency be on the order of (10 GHz), with a sensitivity required for that frequency on the order of $10^{-30}$ δm/m. Once this is obtained, the evolution of cosmological structure can be investigated properly, with the following as targets of opportunity for smart applications of HFGW detectors.

## Time for the headache pills. Not everyone buys dark energy . I.e. Controversies of DM/ DE applications to cosmology. How HFGW may help resolve them.

The following is meant as a travelogue as to current problems in cosmology which will require significant revision of our models. Exhibit A as to what to consider is **The cosmic void hypothesis'. See Timothy Clifton, Pedro G. Ferreira and Kate Land . I.e. Clifton raises the following question- can HFGW and detectors permit cosmologist to get to the bottom of this ? "Solving Einstein's equations for an averaged matter distribution is NOT the same as solving for the real matter distribution and then averaging the resultant geometry"**

("We average, then solve when in effect we should solve, then average") . **New cosmic map (galaxies) reveals colossal structures**

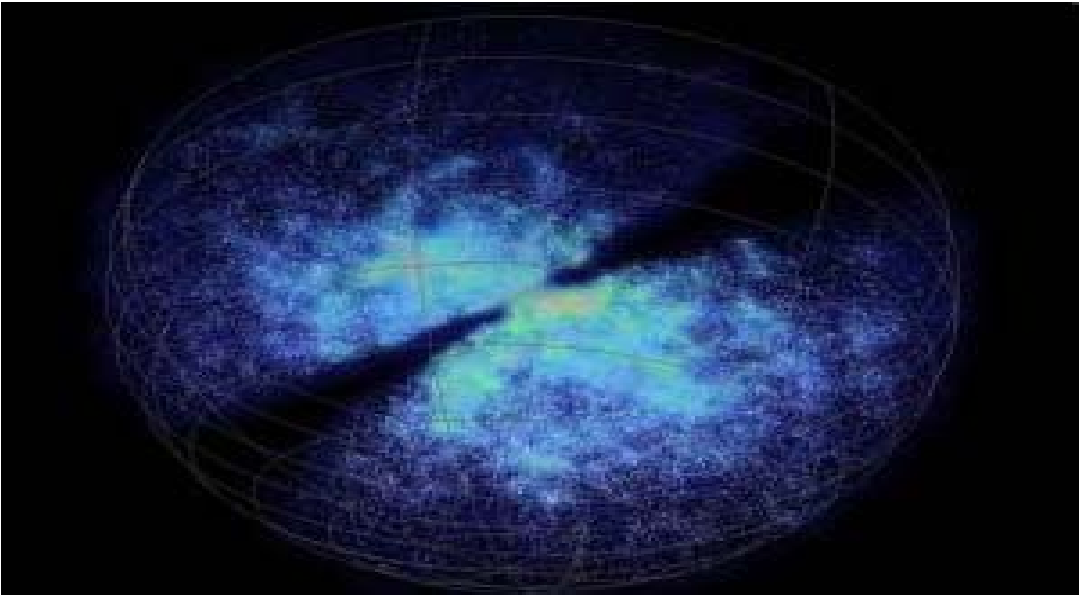

**Figure 2, with huge gaps in Baryonic matter distribution of galaxies.**

The cosmic void hypothesis is extremely important since The new survey mapped the positions of more than 100,000 galaxies. The black strips are areas the survey did not cover because matter in our own galaxy blocked the view."

 "Enormous cosmic voids and giant concentrations of matter have been observed in a new galaxy survey, one of the biggest completed so far. One of the voids is so large that it is difficult to explain where it came from."

- **If there EVER was a good reason to use HFGW as a way to test for matter distribution, THIS LAST figure, figure 2, MAKES THE CASE FOR IT.**

- **What is at stake: The contrast between the VOID picture of expansion, and DE is so stark that this is a guaranteed NOBEL PRIZE in the offering for the first person to falsify either DE, or the VOID hypothesis**

Next, let us look at a recently emerging conundrum of DM feeding into the structure of new galaxies and their far earlier than expected development, i.e. 5 billion years after the big bang. **Galaxy formation issues…. Hierarchical Galaxy Formation theory at a glance usually proceeds as follows. I.e.**

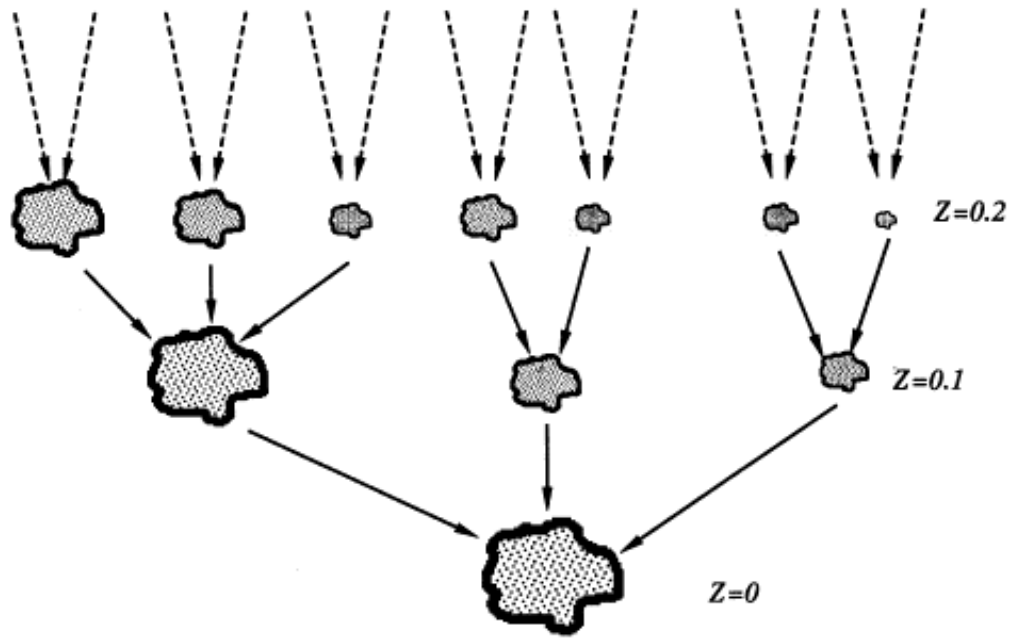

Figure 1. A schematic representation of a halo merging history 'tree'.

**Figure 3. I.e. how we obtain from the 'bottom up' development of galactic super structure.**

What is actually observed, contradicts this halo emerging history 'tree', i.e. Although this 'story' for DM seems to be well established. i.e. Just ONE little problem: DM appears to be fattening up young galaxies, allowing for far-earlier-than-expected creation of early galaxies. **"A clutch of massive galaxies that seem to be almost fully-formed just 5 billion years after the big bang challenge models that suggest galaxies can only form slowly. Tendrils of dark matter that fed the young galaxies on gas could be to blame (NASA/CXC/ESO/P Rosati et al)"**

Source: http://www.newscientist.com/article/dn16912-overweight-galaxies-forcefed-by-dark-matter-tendrils.html. And this may be the result

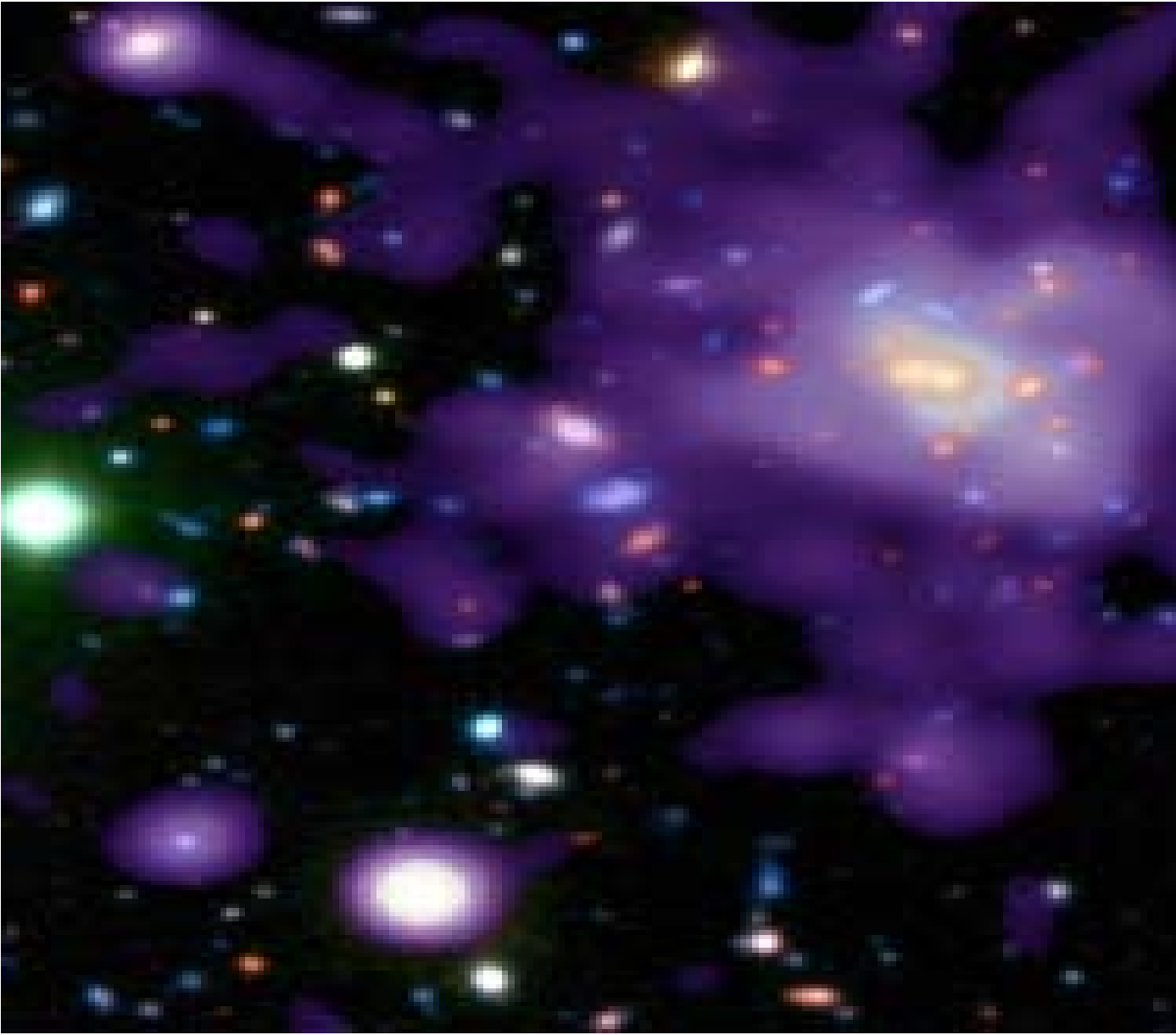

**Figure 4**. Early universe formation of galaxies via DM tendrils. Flatly contradicts Figure 3

## Conclusion.

Let us first reference what can be done with further developments in deformation quantization and its applications to gravitational physics. The most note worthy I have seen centers upon .grassman algebras and deformation quantization of fermionic fields. I.e. Galaviz (2007) showed that one can obtain a Dirac propagator from classical versions of Fermionic fields, and this was a way to obtain minimum quantization conditions for initially classical versions of fermionic fields as due to alterations of algebraic structures, in sutiable ways. One of the aspects of early universe topology we need to consider is how to introduce quantization in curved space time geometries. , and this is a problem which would , among other things permit a curved space treatment of $\Psi \sim [R/R_{eq}]^{3/2}$. I.e. as R gets of the order of $R \sim \vartheta(l_P)$, say that the spatial geometry of early universe expansion is within a few orders of magnitude of Planck length, then how can we recover a field theory quantization condition for $\Psi \sim [R/R_{eq}]^{3/2}$ in terms of path integrals. We claim that deformation quantization, if applied successfully will eventually lead to a great refinement of the above Wheeler De Witt wave functional value, as well as allow a more through match up of a time independent solution of the Wheeler De Witt equation, as given in **Appendix I**, with the more subtle pseudo time dependent evolution of the wave functional as given in **Appendix II.** I .e. the linkage between time independent treatments of the wave functional of the universe, with what Lawrence Crowell wrote up in 2005, will be made more explicit. This will, in addition allow us to understand better how graviton production in relic conditions may add to entropy, as well as how to link the number of gravitons, say $10^{12}$ gravitons per photon, as information as a way to preserve the continuity of $\hbar$ values from a prior universe to the present universe. The author claims that in order to do this rigorously, that use of the material in Gutt, and Waldmann ( ' Deformation of the Poisson bracket on a sympletic manifold' ) as of 2006 will be necessary, especially to recover quantization of severely curved space time conditions which add more detail to $\Psi \sim [R/R_{eq}]^{3/2}$. Having said this, it is now important to consider what can be said about how relic gravitons/ information can pass through minimum vales of $R \sim \vartheta(l_P)$.

We shall reference what the AW. Beckwith (2008) presented in 2008 STAIF, which we think still has current validity for reasons we will elucidate upon in this document. We use a power law relationship first presented by Fontana (2005), who used Park's earlier (1955) derivation: when $E_{eff} \equiv \langle n(\omega) \rangle \cdot \omega \equiv \omega_{eff}$

$$P(power) = 2 \cdot \frac{m_{graviton}^2 \cdot \widehat{L}^4 \cdot \omega_{net}^6}{45 \cdot (c^5 \cdot G)} \qquad (34)$$

This expression of power should be compared with the one presented by Massimo Giovannini (2008) on averaging of the energy-momentum pseudo tensor to get his version of a gravitational power energy density expression, namely

$$\overline{\rho}_{GW}^{(3)}(\tau,\tau_0) \cong \frac{27}{256 \cdot \pi^2} H^2 \cdot \left(\frac{H}{M}\right)^2 \cdot \left[1 + \vartheta \cdot \left(\frac{H^4}{M^4}\right)\right] \qquad (35)$$

Giovannini states that should the mass scale be picked such that $M \sim m_{Planck} \gg m_{graviton}$, that there are doubts that we could even have inflation. However, it is clear that gravitational wave density is faint, even if we make the approximation that $H \equiv \frac{\dot{a}}{a} \cong \frac{m\phi}{\sqrt{6}}$ as stated by Linde (2008), where we are following $\dot{\phi} = -m\sqrt{2/3}$ in evolution, so we have to use different procedures to come up with relic gravitational wave detection schemes to get quantifiable experimental measurements so we can start predicting relic gravitational waves. This is especially true if we make use of the following formula for gravitational radiation, as given by L. Kofman (2008), with $M = V^{1/4}$ as the energy scale, with a stated initial inflationary potential V. This leads to an initial approximation of the emission frequency, using present-day gravitational wave detectors.

$$f \cong \frac{(M = V^{1/4})}{10^7 \, GeV} Hz \qquad (36)$$

For us, getting to the bottom of what was listed, especially in the DM/ DE conundrums, probably by judicious application of the Li – Baker detector, if managed properly, may initiate an exciting era of new cosmology research. This, by necessity will place a premium upon use of a detector system probably along the following lines.

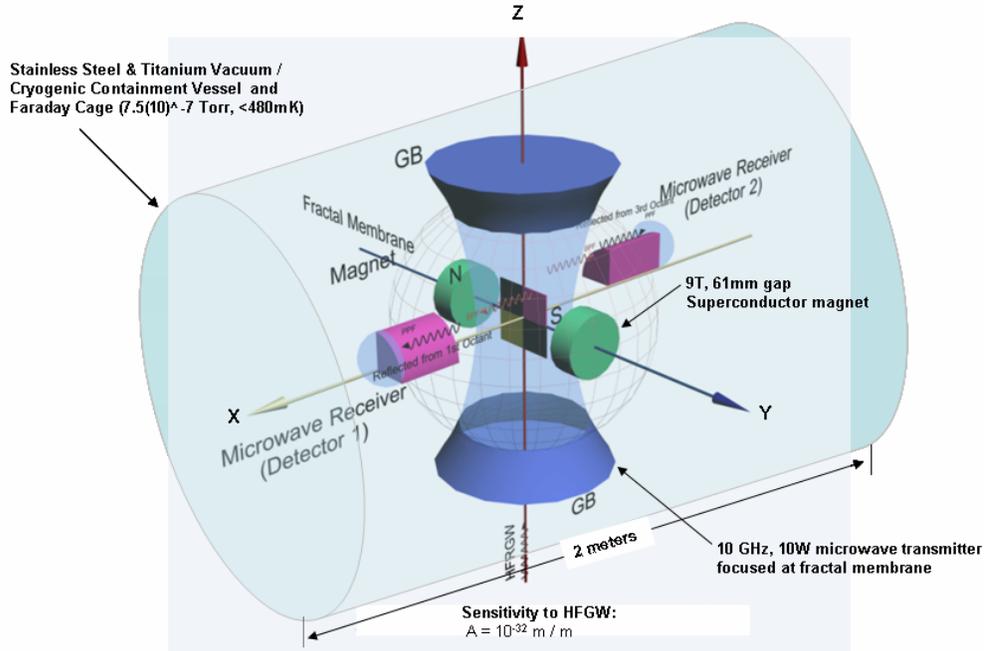

**Figure 5.** A practical HFGW detector as presented / designed by Dr. Li Fangyu and Dr. Baker

## Appendix 1: Linking the thin shell approximation, Weyl quantization , and the Wheeler De Witt equation

This is a re capitulation of what is written by S. Capoziello, et al (2000) for physical review A, which is assuming a generally spherically symmetric line element . The upshot is that we obtain a dynamical evolution equation, similar in part to the Wheeler De Witt equation which can be quantified as $H|\Psi\rangle = 0$ Which in turn will lead to, with qualifications, for thin shell approximations $|x| << 1$.

$$\Psi'' + a^2 x^4 \Psi = 0 \qquad (1)$$

so that $Z_{1/6}$ is a spherical Bessel equation for which we can write

$$\Psi \equiv \sqrt{x} Z_{1/6}\left(\frac{a}{3} x^3\right) \sim x^{2/3} \qquad (2)$$

Similarly, $|x| >> 1$ leads to

$$\Psi \equiv \sqrt{x} Z_{1/6}\left(\frac{a}{3 \cdot \sqrt{2}} x^3\right) \qquad (3)$$

Also, when $x \cong 1$

$$\Psi \equiv \left[\sqrt{2a^2 \cdot (x-1)}\right]^3 Z_{-3/4}\left(\frac{8}{3} \cdot a \cdot (x-1)^{3/2}\right) \quad\quad\quad (4)$$

Realistically, in terms of applications, we will be considering very small $x$ values, consistent with conditions near a singularity/ worm hole bridge between a prior to our present universe. This is for $x \equiv R/R_{equilibrium}$

## Appendix II. How to obtain worm hole bridge between two universes, via the Wheeler De Witt equation : i.e. forming Crowell's time dependent Wheeler-De-Witt equation, and its links to Wormholes

.This will be to show some things about the wormhole we assert the instanton traverses en route to our present universe. This is the Wheeler-De-Witt equation with pseudo time component added. From Crowell

$$-\frac{1}{\eta r} \frac{\partial^2 \Psi}{\partial r^2} + \frac{1}{\eta r^2} \cdot \frac{\partial \Psi}{\partial r} + rR^{(3)}\Psi = \left(r\eta\phi - r\ddot{\phi}\right) \cdot \Psi \quad\quad\quad (1)$$

This has when we do it $\phi \approx \cos(\omega \cdot t)$, and frequently $R^{(3)} \approx$ constant, so then we can consider

$$\phi \cong \int_0^\infty d\omega \left[a(\omega) \cdot e^{ik_\varpi x^\mu} - a^+(\omega) \cdot e^{-ik_\varpi x^\mu}\right] \quad\quad\quad (2)$$

In order to do this, we can write out the following for the solutions to Eqn (1) above.

$$C_1 = \eta^2 \cdot \left(4 \cdot \sqrt{\pi} \cdot \frac{t}{2\omega^5} \cdot J_1(\omega \cdot r) + \frac{4}{\omega^5} \cdot \sin(\omega \cdot r) + (\omega \cdot r) \cdot \cos(\omega \cdot r)\right)$$
$$+ \frac{15}{\omega^5} \cos(\omega \cdot r) - \frac{6}{\omega^5} Si(\omega \cdot r) \quad\quad\quad (3)$$

And

$$C_2 = \frac{3}{2 \cdot \omega^4} \cdot (1 - \cos(\omega \cdot r)) - 4e^{-\omega \cdot r} + \frac{6}{\omega^4} \cdot Ci(\omega \cdot r) \quad\quad\quad (4)$$

This is where $Si(\omega \cdot r)$ and $Ci(\omega \cdot r)$ refer to integrals of the form $\int_{-\infty}^x \frac{\sin(x')}{x'} dx'$ and $\int_{-\infty}^x \frac{\cos(x')}{x'} dx'$. It so happens that this is for forming the wave functional that permits an instanton to form. Next, we should consider whether or not the instanton so formed is stable under evolution of space-time leading up to inflation.

To model this, we use results from Crowell (2005) on quantum fluctuations in space-time, which gives a model from a pseudo time component version of the Wheeler-De-Witt equation, with use of the Reinssner-Nordstrom metric to help us obtain a solution that passes through a thin shell separating two space-times. The radius of the shell $r_0(t)$ separating the two space-times is of length $l_P$ in approximate magnitude, leading to a domination of the time component for the Reissner – Nordstrom metric

$$dS^2 = -F(r) \cdot dt^2 + \frac{dr^2}{F(r)} + d\Omega^2 \quad\quad\quad (5)$$

This has:

$$F(r) = 1 - \frac{2M}{r} + \frac{Q^2}{r^2} - \frac{\Lambda}{3} \cdot r^2 \xrightarrow[T \to 10^{32} Kelvin \sim \infty]{} -\frac{\Lambda}{3} \cdot (r = l_P)^2 \quad (6)$$

This assumes that the cosmological vacuum energy parameter has a temperature dependence as outlined by Park (2003), leading to

$$\frac{\partial F}{\partial r} \sim -2 \cdot \frac{\Lambda}{3} \cdot (r \approx l_P) \equiv \eta(T) \cdot (r \approx l_P) \quad (7)$$

as a wave functional solution to a Wheeler-De-Witt equation bridging two space-times. This solution is similar to that being made between these two space-times with "instantaneous" transfer of thermal heat, as given by Crowell (2005)

$$\Psi(T) \propto -A \cdot \{\eta^2 \cdot C_1\} + A \cdot \eta \cdot \omega^2 \cdot C_2 \quad (8)$$

This has $C_1 = C_1(\omega, t, r)$ as a pseudo cyclic and evolving function in terms of frequency, time, and spatial function. This also applies to the second cyclical wave function $C_2 = C_2(\omega, t, r)$, where we have $C_1 =$ Eqn (3) above, and $C_2 =$ Eqn. (4) above. Eqn. (8) is an approximate solution to the pseudo time dependent Wheeler-De-Witt equation. The advantage of Eqn. (8) is that it represents to good first approximation of gravitational squeezing of the vacuum state.